\documentclass[12pt]{iopart}
\usepackage{iopams}
\usepackage{epsfig}   
\usepackage{graphics}
\newcommand{\apj}{Astrophys. J.}

\newcommand{\apjs}{Astrophys. J. Suppl.}
\newcommand{\aj}{Astron. J.}
\newcommand{\aap}{Astron. Astrophys.}
\newcommand{\prd}{Phys. Rev. D}
\newcommand{\mnras}{Mon. Not. Roy. Astron. Soc.}
\begin{document} 

\title[Supernovae as seen by off-center observers in a local void]
{Supernovae as seen by off-center observers in a local void}

\author{Michael Blomqvist$^1$ and Edvard M\"ortsell$^2$}

\address{$^1$ The Oskar Klein Centre for Cosmoparticle Physics, Department of Astronomy, \\
Stockholm University, AlbaNova University Center \\ S--106 91
Stockholm, Sweden}
\address{$^2$ The Oskar Klein Centre for Cosmoparticle Physics, Department of Physics, \\
 Stockholm University, AlbaNova University Center \\ S--106 91
 Stockholm, Sweden}
\ead{\mailto{michaelb@astro.su.se}, \mailto{edvard@fysik.su.se}}

\begin{abstract}
Inhomogeneous universe models have been proposed as an alternative
explanation for the apparent acceleration of the cosmic expansion that
does not require dark energy. In the simplest class of inhomogeneous
models, we live within a large, spherically symmetric void. Several
studies have shown that such a model can be made consistent with many
observations, in particular the redshift--luminosity distance relation
for type Ia supernovae, provided that the void is of Gpc size and that
we live close to the center. Such a scenario challenges the Copernican
principle that we do not occupy a special place in the universe. We
use the first-year Sloan Digital Sky Survey-II supernova search data
set as well as the Constitution supernova data set to put constraints
on the observer position in void models, using the fact that
off-center observers will observe an anisotropic universe.
We first show that a spherically symmetric void can give good fits to
the supernova data for an on-center observer, but that the two data
sets prefer very different voids. We then continue to show
that the observer can be displaced at least fifteen percent of
the void scale radius from the center and still give an acceptable fit to the
supernova data. When combined with the observed dipole anisotropy of
the cosmic microwave background however, we find that the data
compells the observer to be located within about one percent of the
void scale radius. Based on these results, we conclude that
considerable fine-tuning of our position within the void is needed to
fit the supernova data, strongly disfavouring the model from a
Copernican principle point of view.
\end{abstract}

\noindent{\it Keywords}: dark energy theory, supernova type Ia

\section{Introduction}
The discovery of the dimming of distant type Ia supernovae (SNe~Ia)~\cite{1998AJ....116.1009R,1999ApJ...517..565P} constituted the first
impactful evidence that the expansion of the universe is in a phase of
acceleration. This picture has been corroborated by several
independent probes, including measurements of the cosmic microwave
background (CMB) anisotropies~\cite{2009ApJS..180..330K} and baryon
acoustic oscillations (BAO)~\cite{2005ApJ...633..560E,2007MNRAS.381.1053P}. Under the assumption
that the universe is homogeneous and isotropic, the apparent late-time
acceleration is usually attributed to a mysterious energy component
with negative pressure -- dark energy -- the nature of which is still
unknown.

In recent years, inhomogeneous universe models of varying degrees of
complexity have been the focus of much attention as an alternative
explanation of the apparent acceleration that doesn't require dark
energy. The common base in these models is that the assumption of
homogeneity and isotropy of the universe is an oversimplification and
that it is the presence of inhomogeneities in the distribution of
matter that gives rise to the apparent acceleration. Examples of
inhomogeneous models include the Swiss-cheese models, which consider
light propagation through a universe full of empty regions, and
backreaction models, where cosmological perturbation terms are
included in the overall dynamics of the universe (see~\cite{2007astro.ph..2416C} for a review of inhomogeneous models).

In the simplest class of inhomogeneous models, we live within a large,
spherically symmetric local void described by the
Lema\^{i}tre-Tolman-Bondi (LTB) metric~\cite{1933ASSB...53...51L,1997GReGr..29..641L,1934PNAS...20..169T,1947MNRAS.107..410B}. 
The LTB model is an exact solution of Einstein's equations containing
dust only. In contrast to homogeneous universe models, the Hubble
expansion rate and the matter density parameter depend on both time
and the radial coordinate in LTB models. The apparent acceleration is
caused by the radial dependence, such that our local underdense region
has a larger Hubble parameter than the surrounding homogeneous
universe.

The LTB models violate the Copernican principle by placing the
observer in a special place in the universe. Due to the observed
near-perfect isotropy of the CMB, the observer is generally assumed to
be located very close to the center of the void. Another challenge for
the model is that the void must be of Gpc size in order to fit the
SN~Ia data, but such large voids are extremely improbable in standard
models of structure formation~\cite{2008arXiv0807.4508H}. The LTB
models thus appear to be unrealistic, but based on observations they
have not yet been ruled out. Several studies have shown that with an
appropriately chosen void profile, the LTB models can be made to fit
data from SNe~Ia, CMB and BAO~\cite{2007JCAP...02..019E,2008PhRvL.101m1302C,2006PhRvD..73h3519A,2008JCAP...04..003G,
2009JCAP...02..020B,2008arXiv0810.4939G,2008PhRvL.101y1303Z,2009arXiv0909.1479F}. Further
constraints come from considering spectral distortions of the CMB~\cite{2008PhRvL.100s1302C} and the kinematic Sunyaev-Zeldovich effect~\cite{2008JCAP...09..016G}.

In the most general scenario, the observer can be located anywhere
inside the void, in which case observers living off-center will see an
anisotropic universe. There are several ways to test the level of
anisotropy in the data to put constraints on the position of an
off-center observer. So far, the strongest constraint comes from the
dipole anisotropy of the CMB. The observed CMB dipole restricts a
static observer to be located within a couple of percent of the void
size~\cite{2006PhRvD..74j3520A}. The caveat is -- although it requires a certain 
amount of fine-tuning -- that a peculiar velocity of the observer
directed towards the center can cancel the effect of the off-center
position so that large displacements are allowed. For SNe~Ia,
off-center observers will see an anisotropic relation between the
luminosity distance and the redshift. The SN~Ia data provide
independent constraints on the observer position that complement those
imposed by the CMB~\cite{2007PhRvD..75b3506A}. Another interesting possibility is to look
for cosmic parallax, i.e., a time variation in the angular separation
between distant sources induced by the anisotropic expansion. Planned
space-based astrometric missions (such as Gaia) could measure the
positions of a huge number of quasars over time with outstanding
accuracy and thereby establish constraints on the observer position
that rival those set by the CMB~\cite{2009PhRvL.102o1302Q}.

This paper deals with the contraints on the observer position coming
from SNe~Ia. The problem has been addressed previously by Alnes \&
Amarzguioui~\cite{2007PhRvD..75b3506A} who found that it is possible
to obtain a better fit to the data for an off-center observer. The
constraint on the observer position, however, was found to be fairly
weak, partly due to the low number of SNe~Ia in the data set
employed. Our analysis is similar to theirs, but with a few notable
differences. In addition to using new and larger SN~Ia data sets as
well as a different LTB model, we also investigate whether the model
can fit the SNe~Ia for an off-center observer while simultaneously
accommodate the observed CMB dipole. We find that it is only possible
to obtain a good fit to the data as long as the observer is located
very close to the void center.

We would like to emphasise that our analysis is solely aimed at
probing the impact on the SNe~Ia observations when moving the observer
away from the center of the void. We employ a simple but plausible LTB
model with relatively few parameters to perform this test. We do not
claim, nor believe, that the LTB model investigated here is able to
accommodate all the observations coming from other cosmological probes
as well.

This paper is organised as follows. In section~\ref{model}, we present
the basics of the LTB model and introduce the specific model employed
in this study. In section~\ref{datasets}, we introduce the SNe~Ia data
sets that we use for the analysis. Section~\ref{oncenter} deals with
the case of an on-center observer. In section~\ref{solving}, we
present the differential equations that govern the path and redshift
of the photons seen by an off-center observer. The constraint placed
by SNe~Ia on the observer position is investigated in
section~\ref{position}. In section~\ref{cmb} we look at this
constraint when the CMB dipole is also taken into consideration. The
paper is concluded in section~\ref{conclusions}.

\section{The Lema\^{i}tre-Tolman-Bondi model}\label{model}

A spherically symmetric void can be described mathematically using the
Lema\^{i}tre-Tolman-Bondi metric,
\begin{equation}
ds^2=-dt^2+\frac{A'^2(r,t)}{1-k(r)}dr^2+A^2(r,t)(d\theta^2+\sin^2\theta d\phi^2)\ ,
\end{equation}
where the scale function $A(r,t)$ depends on both time and the radial
coordinate, and $k(r)$ is associated with the spatial curvature. We
use primes and dots to denote partial derivatives with respect to
space and time, respectively.

In LTB models, the Hubble expansion rate $H(r,t)$ for a matter dominated
universe can be written as~\cite{2007JCAP...02..019E}
\begin{equation}\label{Friedmann}
H^2(r,t)=H_{0}^2(r)\Bigg[ \Omega_{\rm M}(r)\Bigg(
\frac{A_0(r)}{A(r,t)} \Bigg)^3+\Omega_{\rm K}(r)\Bigg(
\frac{A_0(r)}{A(r,t)} \Bigg)^2 \Bigg]\ ,
\end{equation}
where the density parameters are related by $\Omega_{\rm M}(r)+\Omega_{\rm K}(r)=1$, and the present-day values $A_0(r)\equiv A(r,t_0)$ and $H_0(r)\equiv H(r,t_0)$. Compared
to the ordinary Friedmann equation of the homogeneous
Friedmann-Robertson-Walker universe, all quantities in
equation~(\ref{Friedmann}) depend on the radial coordinate. The two
arbitrary functions $H_{0}(r)$ and $\Omega_{\rm M}(r)$ define the LTB
models and are as boundary conditions independent, allowing for
inhomogeneities both in the expansion rate and in the matter
density. Equation~(\ref{Friedmann}) can be solved analytically using
an additional parameter $\eta$~\cite{2008JCAP...04..003G},
\begin{equation}\label{Art}
A(r,t)=\frac{\Omega_{\rm M}(r)}{2[1-\Omega_{\rm
M}(r)]}[\cosh(\eta)-1]A_0(r)\ ,
\end{equation}
\begin{equation}\label{H0t}
H_{0}(r)t=\frac{\Omega_{\rm M}(r)}{2[1-\Omega_{\rm
M}(r)]^{3/2}}[\sinh(\eta)-\eta]\ .
\end{equation}

\subsection{Gaussian LTB model}

We consider an LTB model in which the matter density parameter $\Omega_{\rm M}(r)$ takes the form of a Gaussian density fluctuation,
\begin{equation}
\Omega_{\rm M}(r)=\Omega_{\rm out}+(\Omega_{\rm in}-\Omega_{\rm out})e^{-(r/r_{\rm s})^{2}}\ .
\end{equation}
The model has three free parameters, where $\Omega_{\rm in}$ is the
matter density at the center of the void, $\Omega_{\rm out}$ is the
asymptotic value of the matter density and $r_{\rm s}$ is the scale
radius of the density fluctuation. This density profile is, of course,
just a toy model, but it provides the basic requirement of a smooth
transition from the local to the distant matter density, without
introducing too many new parameters.

Our LTB model is more constrained than the general case in the sense
that we impose that the Big Bang occured simultaneously throughout
space by implementing a particular choice of $H_{0}(r)$,
\begin{equation}\label{h0r}
H_{0}(r)=\frac{3H_{0}}{2}\Bigg[ \frac{1}{\Omega_{\rm
K}(r)}-\frac{\Omega_{\rm M}(r)}{\sqrt{\Omega_{\rm K}^{3}(r)}}\sinh
^{-1}{\sqrt{\frac{\Omega_{\rm K}(r)}{\Omega_{\rm M}(r)}}} \Bigg]\ ,
\end{equation}
so that the time since the Big Bang is $t_{\rm
BB}=\frac{2}{3}H_{0}^{-1}$ for all observers irrespective of their
position in space. The model is thus completely specified by only one
free function, the matter density $\Omega_{\rm M}(r)$. Note that the functional form
of $\Omega_{\rm M}(r)$ corresponds to a local underdensity also in the
physical matter density. The density contrast, i.e., the ratio between the local and the asymptotic physical matter density, decreases with time, so that the void grows deeper as time progresses. We have
introduced the pre-factor of 3/2 in equation~(\ref{h0r}) to normalize
the age of the universe to that of the Einstein--de Sitter universe. The
constant $H_0$ sets the scale of the expansion rate and determines the
age of the universe for the model. Note that $H_0$ does not represent
the local expansion rate in this model. We marginalize over $H_0$ in
the SN~Ia fit, so its value is arbitrary in the procedure (as long as
$r_{\rm s}$ is in units of Gpc~$h^{-1}$). However, for the purpose of
establishing an absolute distance scale for the analysis, we choose
$H_0=50$~km~s$^{-1}$~Mpc$^{-1}$\footnote{Since distances depend
linearly on $H_0$, it is easy to convert the quoted values to any
other scale determined by a different $H_0$ value.}. This value gives
an age of the universe $t_{\rm BB}\sim 13$~Gyr and a local expansion
rate $H_{0}(r=0)\sim 65$~km~s$^{-1}$~Mpc$^{-1}$ for the model.

Based on predictions from inflation, it is common practice to impose
the prior of asymptotic flatness, $\Omega_{\rm out}=1$. We will for
the most part of the analysis allow $\Omega_{\rm out}$ to be any
positive value $\le 1$, i.e., the models need not be asymptotically
flat. In principle we also allow for solutions with a local
\textit{over}density. In this way we can better probe what kind of
void the SNe~Ia data prefer and also leave open the
possibility for curvature in the distant universe. We will discuss and
compare how imposing asymptotic flatness affects the results.

\section{Data sets}\label{datasets}

\begin{figure}
\begin{center}
\includegraphics[angle=0,width=.45\textwidth]{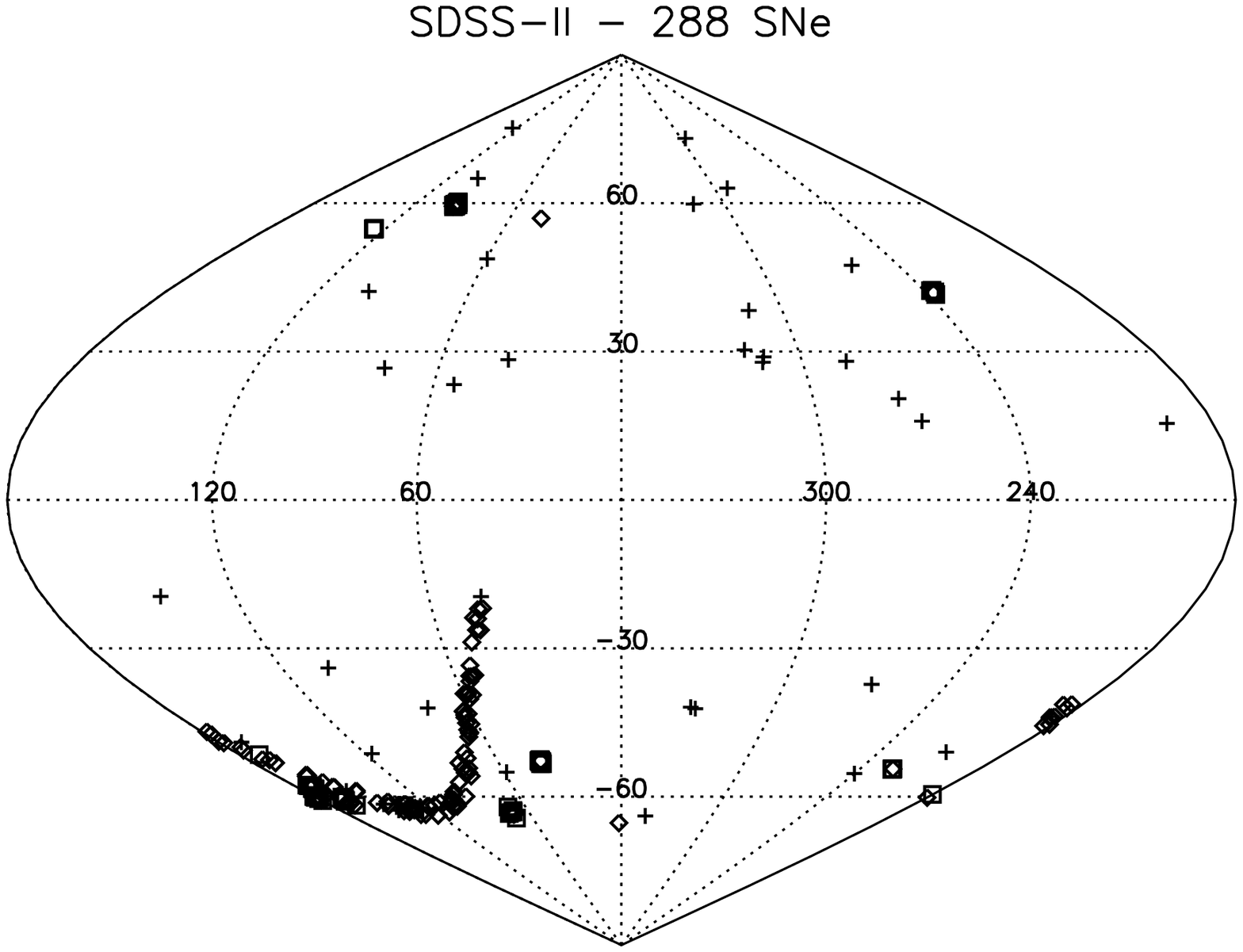}
\includegraphics[angle=0,width=.45\textwidth]{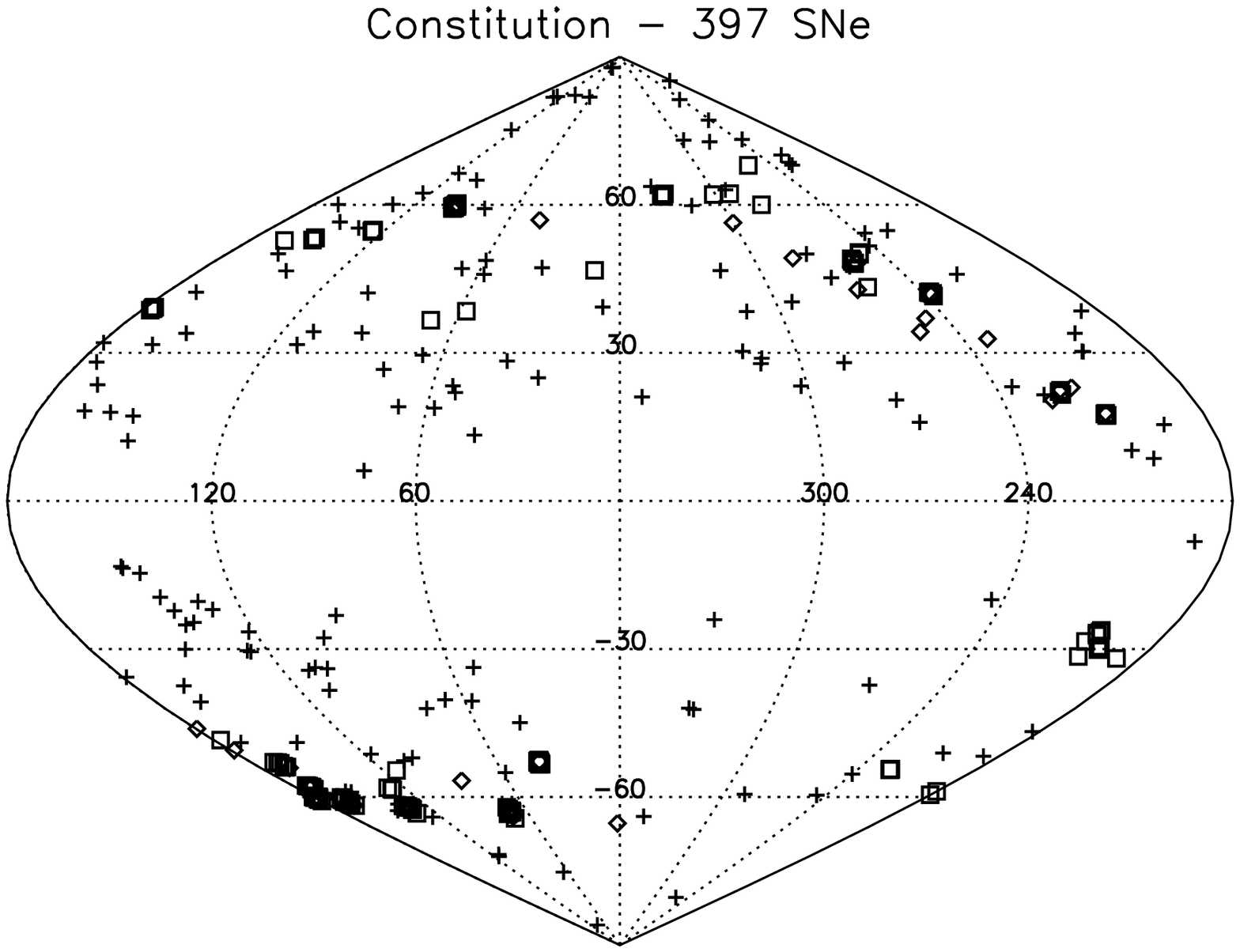}
\caption{\label{fig:sdssmap} Distribution of SNe~Ia in the sky in
galactic coordinates. Left panel: 288 SNe~Ia in the SDSS-II data
set. SNe~Ia with $z<0.1$ are marked with pluses, $0.1<z<0.4$ with
diamonds and $z>0.4$ with squares. Right panel: 397 SNe~Ia in the
Constitution data set, using the same symbols.}
\end{center}
\end{figure}

For the analysis, we employ two recent data sets from the
literature. The first set is the first-year Sloan Digital Sky
Survey-II (SDSS-II) Supernova Search data set presented in Kessler
\textit{et al}~\cite{2009arXiv0908.4274K}. The compiled data set is
based on 103 SNe~Ia at intermediate redshifts discovered by the
SDSS-II, but also includes previously reported SNe~Ia at low and high
redshift from other surveys (nearby~\cite{2007ApJ...659..122J}, ESSENCE~\cite{2007ApJ...666..694W}, SNLS~\cite{2006A&A...447...31A} and HST~\cite{2007ApJ...659...98R}), bringing the total number of SNe~Ia in
the data set to 288. The distance moduli and uncertainties were
obtained using the MLCS2k2 light-curve fitter. The uncertainties
include both the observational and the intrinsic magnitude
scatter. Figure~\ref{fig:sdssmap} (left panel) shows the sky
distribution of the SNe~Ia in the SDSS-II data set. SNe~Ia with
$z<0.1$ are marked with pluses, $0.1<z<0.4$ with diamonds and $z>0.4$
with squares. The data set has a comparatively strong emphasis on
intermediate redshifts with 132 SNe~Ia in this range. The SNe~Ia
discovered by the SDSS-II cover the redshift interval
$z=[0.045,0.421]$ and lie on a thin stripe along the equatorial
plane. Whereas the 41 nearby SNe~Ia are fairly evenly scattered across
the sky, the 115 high $z$ SNe~Ia are confined to small patches. While
this sky distribution is not ideal for investigating anisotropies in
the Hubble diagram, by filling the previously underexplored
intermediate redshift range, the SDSS-II data set is very interesting
for testing LTB models since void sizes of the order of 1~Gpc
correspond to redshifts in this range.

The second data set used in our analysis is the Constitution set
presented in Hicken \textit{et al}~\cite{2009ApJ...700.1097H}. The
data set consists of 397 SNe~Ia and extends the previously available
Union set~\cite{2008ApJ...686..749K} (which includes many SNe~Ia from
ESSENCE, SNLS and HST) by adding 90 SNe~Ia at low reshifts discovered
by the CfA3~\cite{2009ApJ...700..331H}. The distance moduli and
uncertainties were obtained using the SALT light-curve fitter. The
uncertainties include both the observational and the intrinsic
magnitude scatter. Figure~\ref{fig:sdssmap} (right panel) shows the
sky distribution of the SNe~Ia in the Constitution data set. The set
contains 141 SNe~Ia at $z<0.1$ (pluses), which are evenly distributed
across the sky, and 200 SNe~Ia at $z>0.4$ (squares), of which the
majority are confined to small patches. The Constitution set lacks
proper coverage at intermediate redshifts. Only 56 SNe~Ia occupy the
range $0.1<z<0.4$ (diamonds).

Performing the analysis using these two different data sets is an
interesting comparison for a couple of reasons. The difference in the
redshift distributions affects the type of void that the data sets prefer. 
Meanwhile, the larger number of SNe~Ia and
better sky distribution of the Constitution set provide stronger
constraints on the observer position. Another important factor is that the data
sets also differ in the light-curve fitter used to obtain the
distance moduli and uncertainties. Such systematic differences can
have a large impact on the results of any cosmological test using
SNe~Ia~\cite{2009arXiv0908.4274K}.

\section{LTB model for an on-center observer}\label{oncenter}
In this section we look at the situation when the observer is located
at the center of the void and investigate the constraints that the
data sets infer on the parameters of the LTB model. These parameter
values are then used as our starting point in subsequent sections when
the observer is moved off-center.

\subsection{Calculating the luminosity distance for an on-center observer}
An observer located at the center of the void sees a spherically
symmetric universe. Incoming light travels along radial null geodesics
so that the relation between the redshift and the coordinates is given
by a pair of differential equations~\cite{2007JCAP...02..019E},
\begin{equation}\label{dtdz}
\frac{dt}{dz}=-\frac{A'(r,t)}{(1+z)\dot{A}'(r,t)}
\end{equation}
\begin{equation}\label{drdz}
\frac{dr}{dz}=\frac{\sqrt{1-k(r)}}{(1+z)\dot{A}'(r,t)}\ .
\end{equation}
Equations~(\ref{dtdz}) and (\ref{drdz}) determine $t(z)$ and
$r(z)$. Combining these with equations~(\ref{Art}) and (\ref{H0t}), we
can calculate the angular diameter distance measured by an on-center
observer as
\begin{equation}\label{dA}
d_{\rm A}=A[r(z),t(z)]\ .
\end{equation}
The luminosity distance is related to the angular diameter distance
according to
\begin{equation}
d_{\rm L}=(1+z)^2d_{\rm A}\ ,
\end{equation}
and the distance modulus, which is the difference between the apparent
magnitude $m$ and the absolute magnitude $M$, is given by
\begin{equation}
\mu\equiv m-M=5\log _{10}\left(\frac{d_{\rm L}}{1\ \rm Mpc}\right)+25\ .
\end{equation}

\subsection{Results}

\begin{figure}
\begin{center}
\includegraphics[angle=0,width=.48\textwidth]{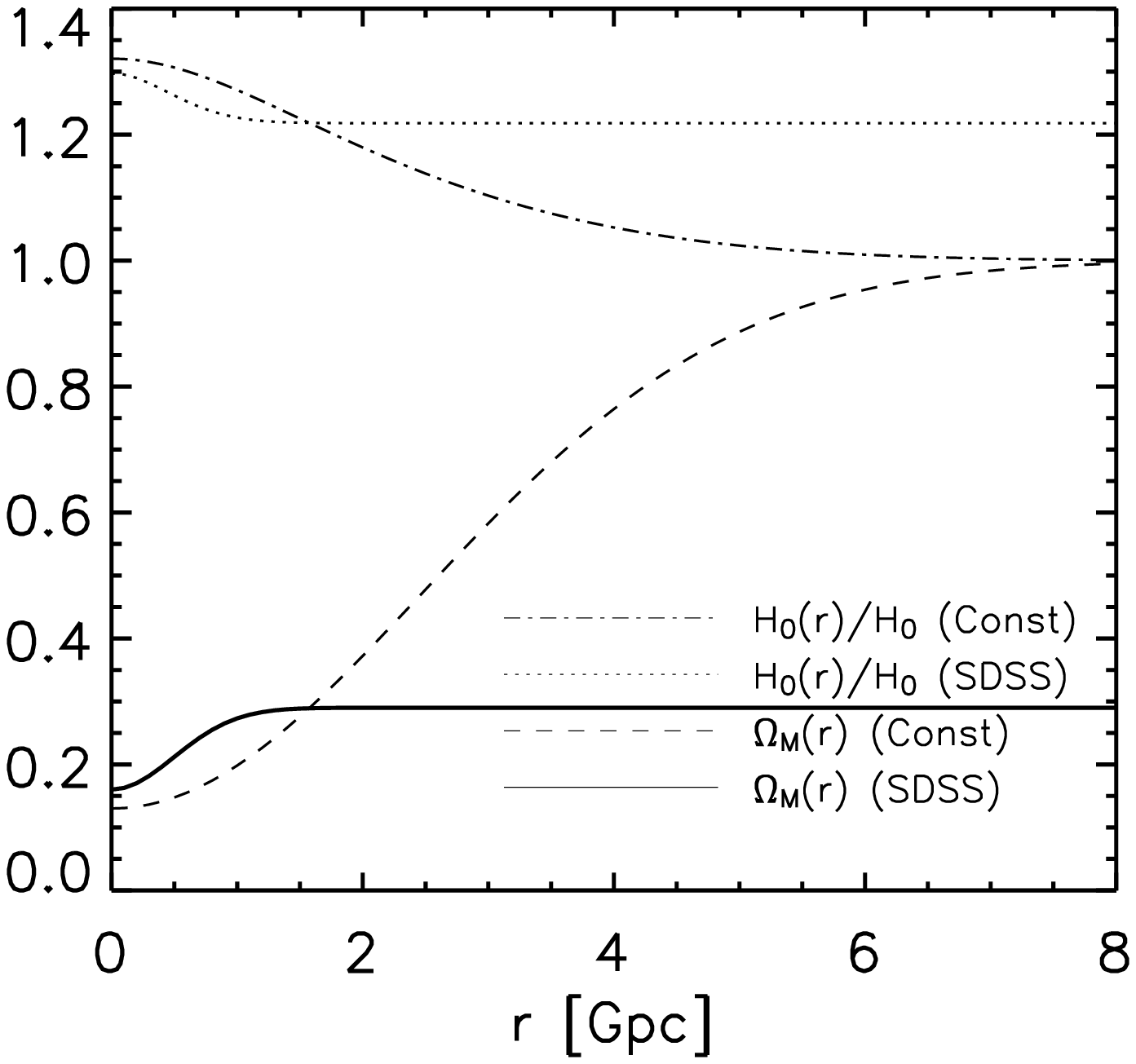}
\includegraphics[angle=0,width=.48\textwidth]{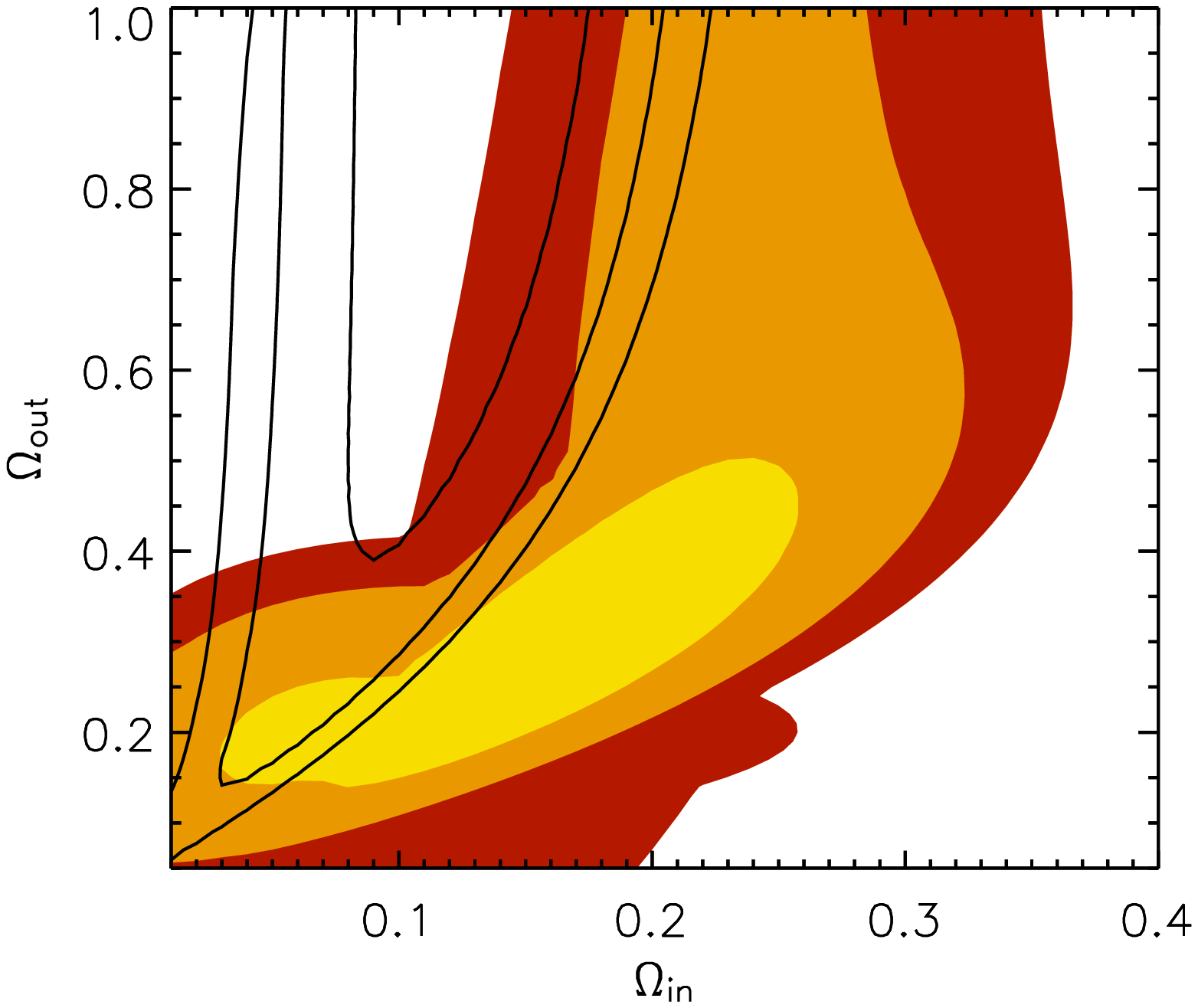}
\includegraphics[angle=0,width=.48\textwidth]{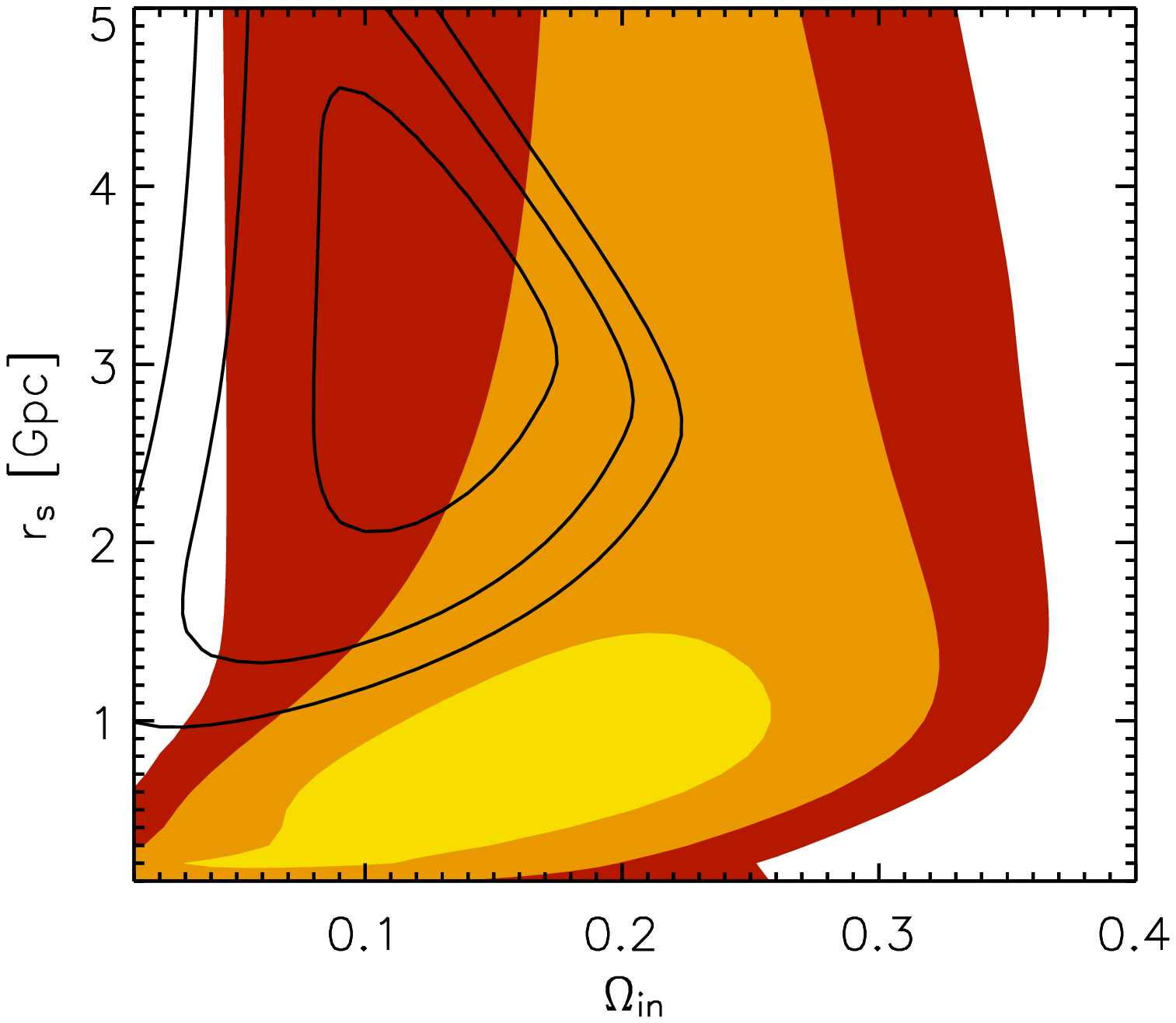}
\includegraphics[angle=0,width=.48\textwidth]{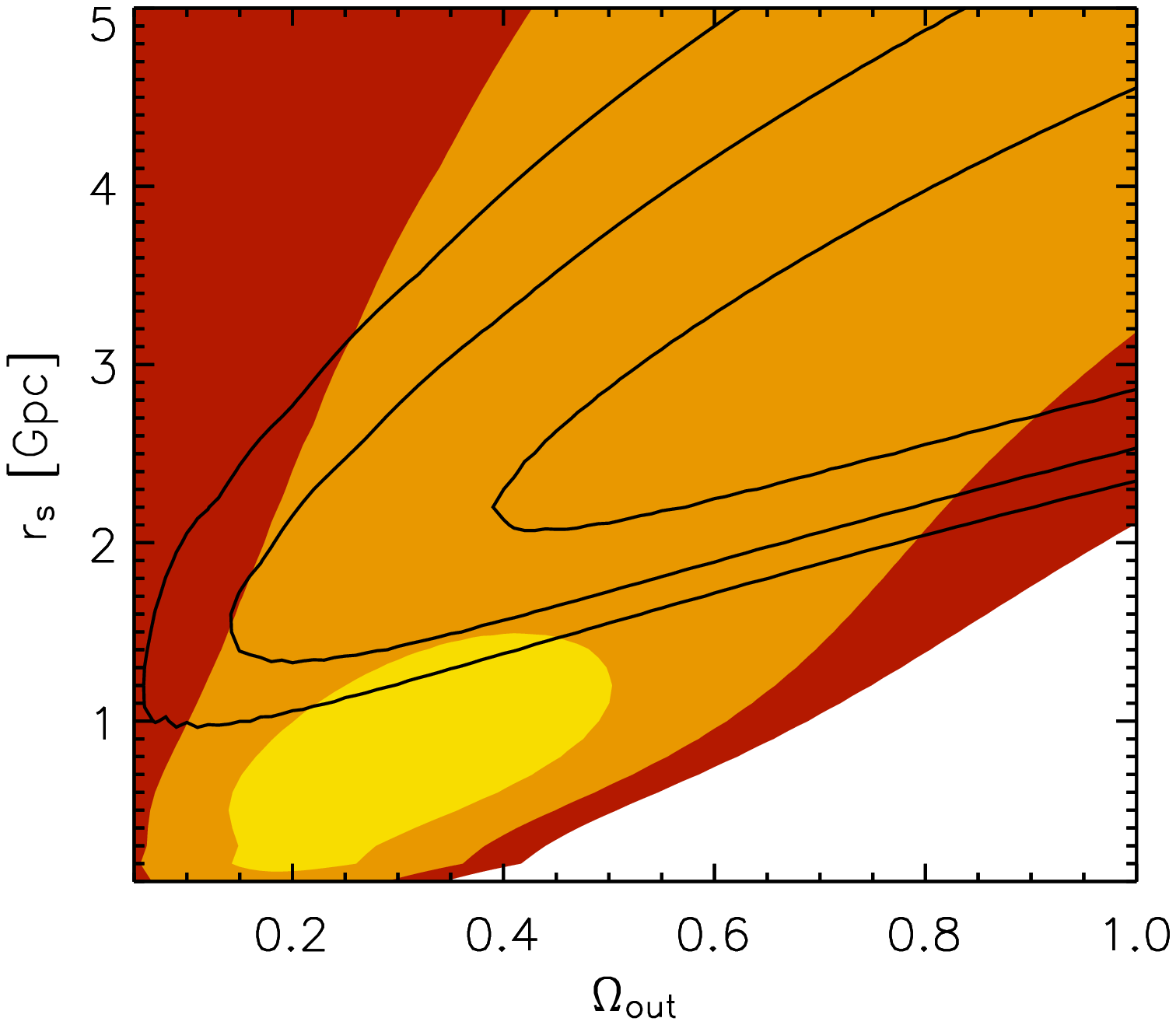}
\caption{\label{fig:LTBmodel} The best fit on-center LTB models. 
The upper left panel shows $\Omega_{\rm M}(r)$ and $H_{0}(r)$ as a
function of the radial coordinate. The other panels show the 68.3, 95
and 99\,\% confidence contours (for two parameters) for the model
parameters $\Omega_{\rm in}$, $\Omega_{\rm out}$ and $r_{\rm s}$. The
contours are given in colour for the SDSS-II data set and with solid
black lines for the Constitution set.}
\end{center}
\end{figure}

The best fit LTB model to the SDSS-II data set for an on-center
observer has the best fit parameters $\Omega_{\rm in}=0.16$,
$\Omega_{\rm out}=0.29$ and $r_{\rm s}=0.7$~Gpc (corresponding to
$z\approx 0.16$) with $\chi^2=229.3$. In comparison, the best fit flat
$\Lambda$CDM model has $\chi^2=231.3$. For the Constitution set the
best fit LTB model has $\Omega_{\rm in}=0.13$, $\Omega_{\rm out}=1$
and $r_{\rm s}=3.5$~Gpc (corresponding to $z\approx 1.02$) with
$\chi^2=461.0$. Here the flat $\Lambda$CDM model gives
$\chi^2=465.5$. Figure~\ref{fig:LTBmodel} shows the results for the
two data sets. The top left panel shows the functions $\Omega_{\rm
M}(r)$ and $H_{0}(r)$ and illustrates well the big difference between
the best fit models. Whereas the SDSS-II data are best fit with a
moderately large void with a small change in the matter density, the
Constitution set prefers an asymptotically flat model with a very
large void. The other panels show the 68.3, 95 and 99\,\% two
parameter confidence contours (corresponding to
$\Delta\chi^2=[2.30,5.99,9.21]$) for the model parameters, with the
SDSS-II contours presented in colour and the Constitution contours
overplotted with solid lines. There is a clear tension between the
data sets, with no overlap of the 68.3\,\% contours. This difference in results is partly a consequence of the different redshift distributions, but also an effect of the light-curve fitter used\footnote{We note that we see tension also in the best fit flat $\Lambda$CDM model, where $\Omega_{\rm m}=0.40\pm0.04$ (68.3\,\% confidence limit) for the SDSS-II set and $\Omega_{\rm m}=0.29\pm0.03$ for the Constitution set.}. If we impose the
prior of asymptotic flatness, $\Omega_{\rm out}=1$, the SDSS-II fit
becomes notably worse, with $\chi^2=234.6$. The local matter density
changes to $\Omega_{\rm in}=0.23$ and the scale radius is pushed up to a
whopping $r_{\rm s}=4.8$~Gpc. For the Constitution set this prior does
not make a difference, since an asymptotically flat model is already
preferred by the fit.

\begin{figure}
\begin{center}
\includegraphics[angle=0,width=.48\textwidth]{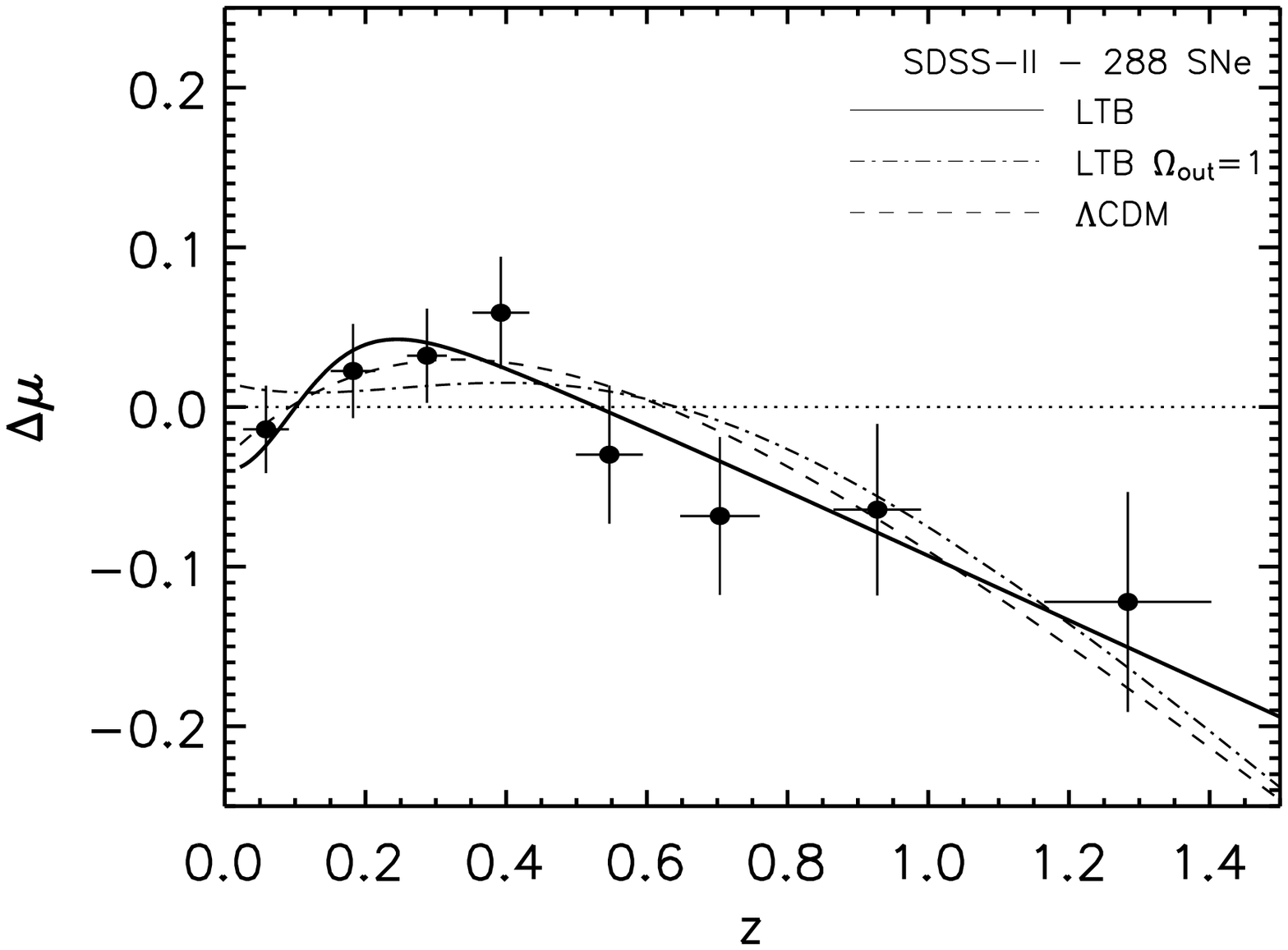}
\includegraphics[angle=0,width=.48\textwidth]{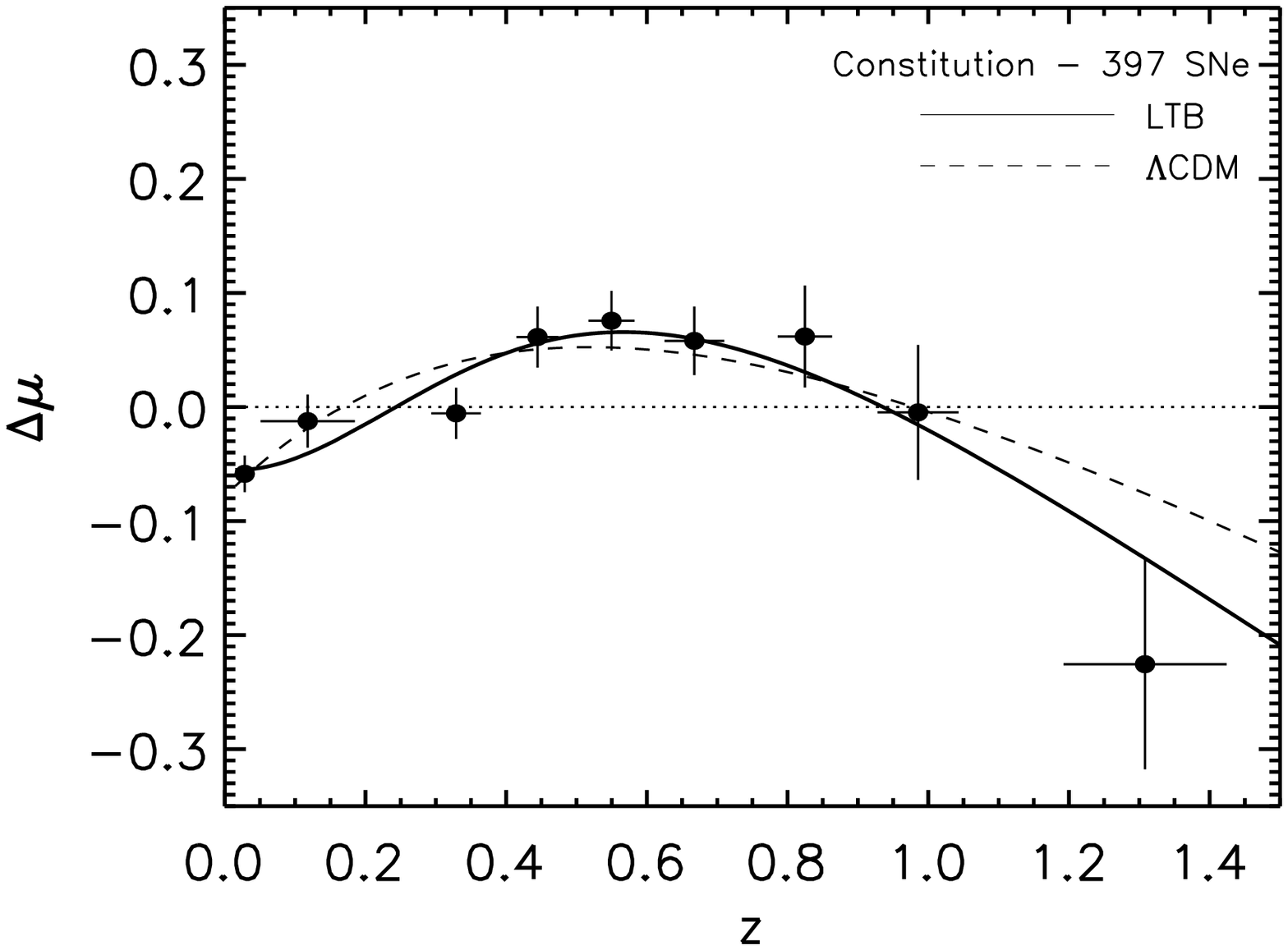}
\caption{\label{fig:hubblediag} Hubble diagrams for the best fit 
on-center models (solid). Distance modulus differences in magnitudes
are shown with respect to an empty universe. The data have been binned
for visualization purposes using $n\Delta z=5$, where $n$ is the
number of SNe~Ia in the bin and $\Delta z$ the redshift range. The
data points are located at the mean value of the redshifts and the
weighted mean of the distance moduli. The redshift error bars show the
standard deviation of the redshifts in the bin, while the distance
modulus error bars give the error of the weighted mean of the distance
moduli within the bin. The best fit flat $\Lambda$CDM models (dashed)
are included for comparison. For the SDSS-II case, we also show the
best fit asymptotically flat ($\Omega_{\rm out}=1$) LTB model
(dot-dashed).}
\end{center}
\end{figure}

The best fits to the data sets are also illustrated in the Hubble
diagrams in Figure~\ref{fig:hubblediag}. For comparison, we have
included the best fit flat $\Lambda$CDM models, and for the SDSS-II
case we have also plotted the best fit asymptotically flat LTB
model. It is quite striking how the data differ between the
panels. The difference in the best fit parameter values for the data sets can be
appreciated fairly intuitively from this figure. For the SDSS-II data
set (left panel), the new SNe~Ia at intermediate redshifts trace out a bump
centered around $z\sim 0.25$, which largely dictates the size of the
void. A smaller void would place the bump at too low redshift, while a
larger void would move the bump to higher redshift and make it less
pronounced. At higher redshifts, i.e., in the regime where
$\Omega_{\rm M}(r)$ is essentially constant, the curve follows close
to a straight line. The low value of $\Omega_{\rm out}=0.29$, ensures
that it doesn't fall off too fast and misses the high-$z$ data
bins. The Constitution set (right panel), on the other hand, contains few
SNe~Ia at intermediate redshifts, but instead features twice as many
SNe~Ia as the SDSS-II between $z=0.4$ and $z=0.9$. These give a very
broad bump centered around $z\sim 0.6$ in the Hubble diagram. The data
is best fit with a very large void together with a high value for
$\Omega_{\rm out}$. A smaller void would place the bump at lower
redshift, while a decrease in the asymptotic matter density would
flatten out the curve.

We also mention that we have investigated the Gaussian LTB model
previously for the case of an on-center observer and ranked it against
the most commonly considered dark energy models~\cite{2009ApJ...703.1374S}. We used the SDSS-II data set but also
included constraints from the cosmic microwave background and the
baryon acoustic oscillations in the analysis. While the LTB model
could provide a fit that was comparable to those of the dark energy
models, it was not preferred from a model selection point of view
since extra parameters are required.

\section{Solving the geodesic equation for off-center observers}\label{solving}
We now wish to calculate the luminosity distance to a SN~Ia as seen by
an observer located off-center in the void. We define the origin of
the coordinate system to be at the void center and choose the $z$-axis
in the direction of the off-center observer. The coordinates of any
space-time point in this frame are the time coordinate $t$, the radial
coordinate $r$, the polar angle $\theta$ and the azimuthal angle
$\phi$. This means that the spatial coordinates of the observer in
this frame are $r=r_{\rm obs}$, $\theta=0$ and a degenerate
$\phi$. Infalling photons hit the observer at time $t_0$, polar angle
$\gamma$ and azimuthal angle $\xi$. We need to trace these photons
back along their trajectories to the source. The redshift of the SN~Ia
refers to what the off-center observer measures, such that $z(r_{\rm
obs})=0$.

The system of differential equations that we need to solve in order to
obtain the coordinates of the SN~Ia as a function of redshift is
derived in \ref{append},
\begin{equation}
\frac{dt}{dz}=-\frac{(1+z)}{q}\ ,
\end{equation}
\begin{equation}
\frac{dr}{dz}=\frac{p}{q}\ ,
\end{equation}
\begin{equation}
\frac{d\theta}{dz}=\frac{J}{qA^2}\ ,
\end{equation}
\begin{equation}
\frac{dp}{dz}=\frac{1}{q}\Bigg[ \frac{(1-k)}{A'}\frac{J^2}{A^3}+\frac{2\dot{A}'}{A'}p(1+z)-
\Bigg( \frac{A''}{A'}+\frac{k'}{2-2k} \Bigg)p^2\Bigg]\ ,
\end{equation}
under the constraint
\begin{equation}
q=\Bigg[ \frac{A'\dot{A}'}{1-k}p^2+ \frac{\dot{A}J^2}{A^3}\Bigg]\ ,
\end{equation}
where $p\equiv dr/d\lambda$, with $\lambda$ being an affine parameter
along the geodesic. The constant angular momentum $J$ can be
calculated as~\cite{2006PhRvD..74j3520A}
\begin{equation}
J=A(r_{0},t_{0})\sin \gamma \ .
\end{equation}
The initial conditions used for the off-center observer are
\begin{equation}
t_{0}=t_{\rm BB}
\end{equation}
\begin{equation}
r_{0}=r_{\rm obs}
\end{equation}
\begin{equation}
\theta_{0}=0
\end{equation}
\begin{equation}
p_{0}=\frac{\sqrt{1-k(r_{0})}}{A'(r_{0},t_{0})}\cos \gamma \ .
\end{equation}

The SN~Ia positions in the sky are in the equatorial coordinate
system given by the right ascension $\alpha$ and declination
$\delta$. We need to relate these angles to the coordinate system of
the void. The polar angle $\gamma$ seen by the off-center observer can
be obtained as
\begin{equation}
\cos \gamma =-\sin \beta \sin(90^\circ-\delta)\cos(\alpha+\sigma)+
\cos \beta \cos(90^\circ-\delta)\ ,
\end{equation}
where $\beta$ is the inclination angle between the $z$-axis and the
celestial equator's normal and $\sigma$ is the angle between the
$x$-axis and the vernal equinox point. The orientation angles $\beta$
and $\sigma$ determine the position of the void center and are free
parameters in the fit. With these definitions, the void center is
located at $\alpha_{\rm c}=360^\circ-\sigma$ and $\delta_{\rm
c}=90^\circ-\beta$ in the equatorial coordinate system. In moving the
observer away from the void center, we have thus introduced three new
free parameters in the fit; the radial displacement $r_{\rm obs}$ and
the orientation angles $\beta$ and $\sigma$.

Finally, the expression for calculating the angular diameter distance
is modified for the case of an off-center observer to~\cite{2007PhRvD..75b3506A},
\begin{equation}
d_{\rm A}^4=\frac{A^4\sin^2\theta}{\sin^2\gamma}\Bigg[ \frac{A'^2}{A^2(1-k)}
\Bigg( \frac{\partial r}{\partial \gamma} \Bigg)^2+\Bigg( \frac{\partial \theta}
{\partial \gamma} \Bigg)^2 \Bigg]\ ,
\end{equation}
where the partial derivatives are obtained numerically in the solution
of the geodesic equations.

\section{Constraining the observer position with SNe~Ia}\label{position}
Off-center observers will see an anisotropic relation between the
luminosity distance and the redshift for the SNe~Ia. This means that a
standard candle with the same redshift but in different directions in
the sky will have different observed magnitudes. The isotropy of the
data can be used to establish constraints on the observer position
inside the void. In this section, we will investigate how far from the
center the observer can be located.

\subsection{Maximum anisotropy}

\begin{figure}
\begin{center}
\includegraphics[angle=0,width=.48\textwidth]{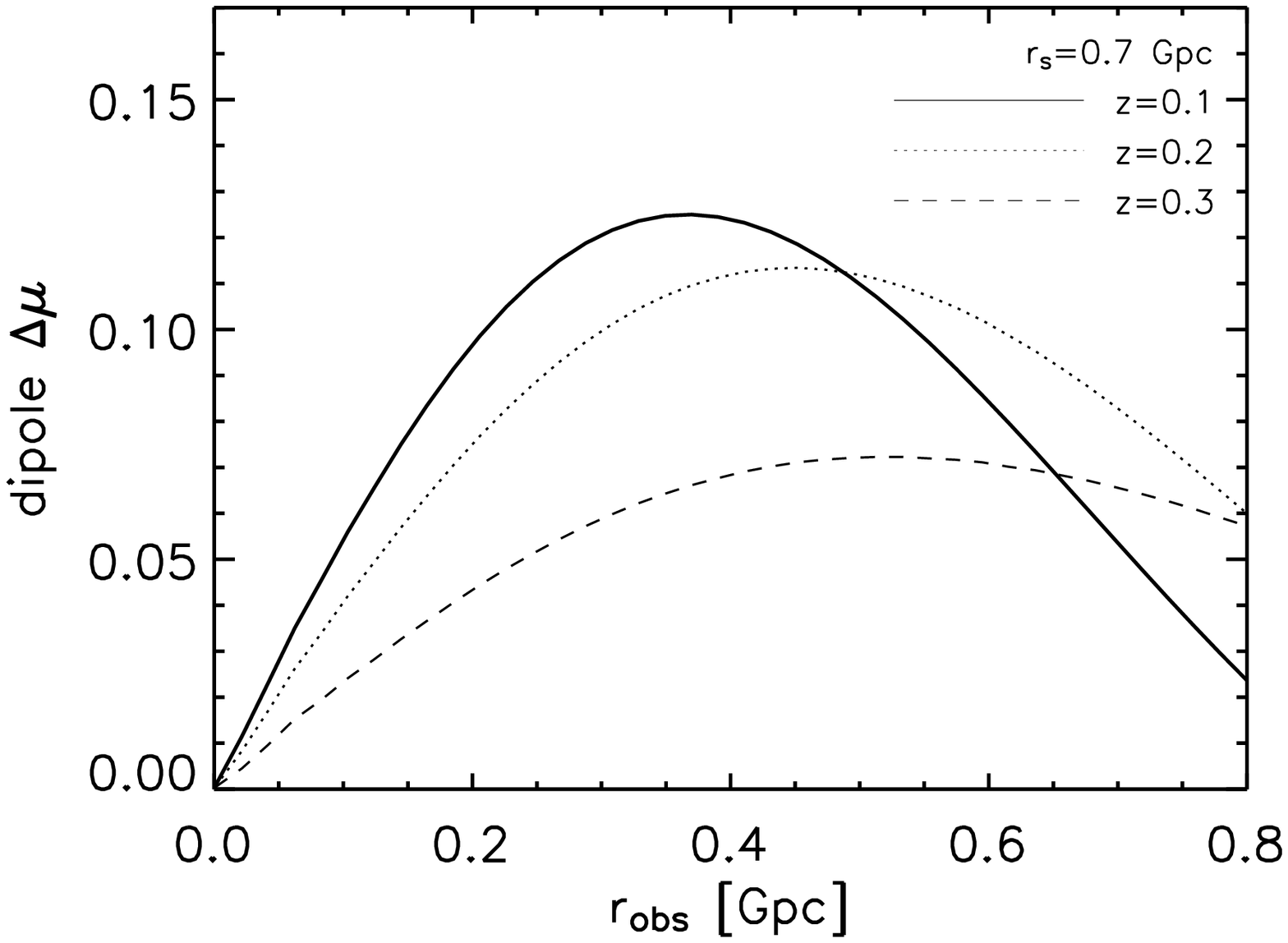}
\includegraphics[angle=0,width=.48\textwidth]{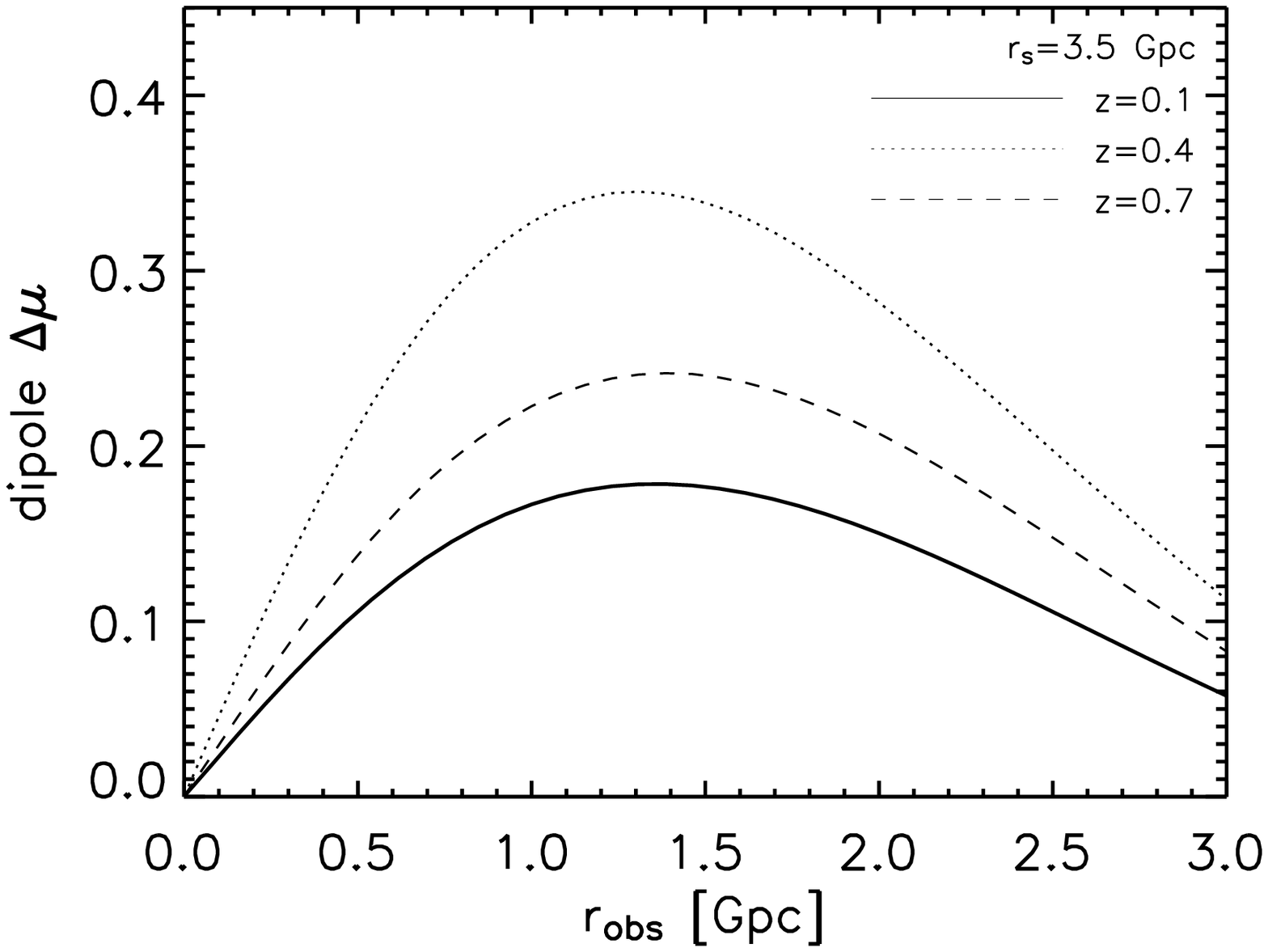}
\caption{\label{fig:dipole} Magnitude dipole induced by moving the 
observer away from the void center in the best fit on-center
models. The curves show the difference in magnitude for two SNe~Ia
with the same redshift but in opposite directions in the sky. Left
panel: A void with scale radius $r_{\rm s}=0.7$~Gpc ($z\approx 0.18$),
preferred by the SDSS-II data set. Right panel: A void with scale radius
$r_{\rm s}=3.5$~Gpc ($z\approx 1.02$), preferred by the Constitution
data set.}
\end{center}
\end{figure}

To get a sense for how big the effect of being situated off-center has
on the SN~Ia observations, we can calculate the maximum anisotropy in
the form of the magnitude dipole. We take two SNe~Ia with the same
redshift but in opposite directions in the sky, aligned with the
off-center observer through the void center. Figure~\ref{fig:dipole}
shows the magnitude dipole, i.e., the difference in magnitude between
the two SNe~Ia for three different redshifts as a function of the
radial displacement of the observer. We have used the best fit
on-center models for the SDSS-II (left panel), where $r_{\rm s}=0.7$~Gpc ($z\approx 0.16$), and the Constitution data set (right panel),
where $r_{\rm s}=3.5$~Gpc ($z\approx 1.02$). 

The behaviour of the curves is easily understood. For on-center
observers, the universe is isotropic and the magnitude dipole
vanishes. For observers located very far from the center, the
magnitude dipole becomes less significant. The curves reach a maximum
at some displacement, depending on the
redshift. Figure~\ref{fig:dipole} demonstrates that SNe~Ia at
different redshifts have different constraining power when determining
the observer position. For the smaller void preferred by the SDSS-II
data, the largest anisotropies are obtained for SNe~Ia at $z=0.1$. The
difference in brightness can be larger than 0.1 magnitudes if the
observer is located a few hundred Mpc from the center. SNe~Ia at
larger redshifts are less affected by the off-center position. SNe~Ia
at low to intermediate redshifts will provide the strongest
constraints on the observer position in this case. In the much larger
void obtained for the Constitution set, SNe~Ia at $z=0.4$ display the
largest anisotropy. An observer located about 1-2 Gpc from the center
would see a magnitude dipole of around 0.3 magnitudes for these
SNe~Ia. For this void, the strongest constraints on the observer
position come from SNe~Ia at intermediate to high redshift. These
conclusions also hold for the SDSS-II fit when we impose a prior of
asymptotic flatness on the model.

\subsection{Results}

\begin{figure}
\begin{center}
\includegraphics[angle=0,width=.48\textwidth]{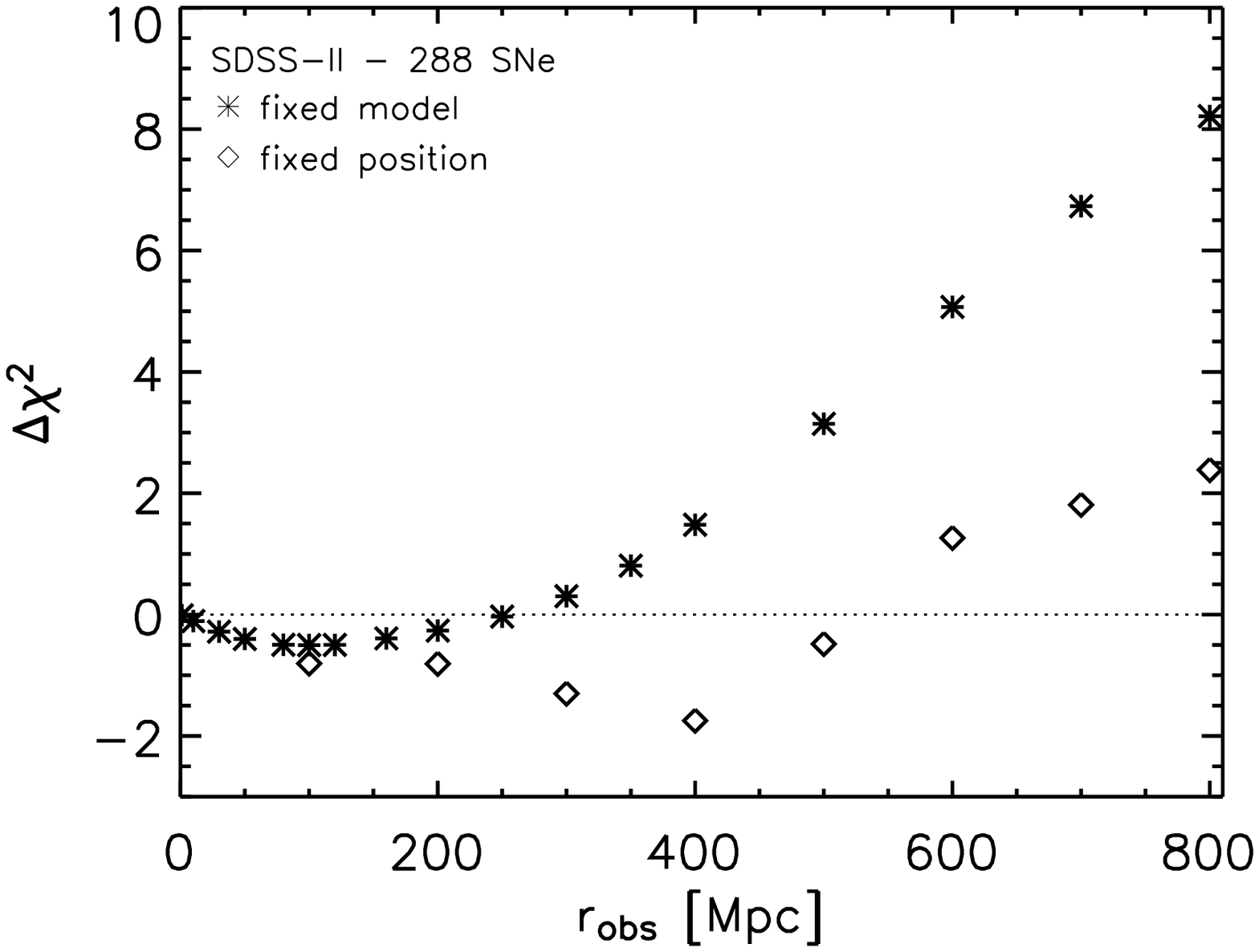}
\includegraphics[angle=0,width=.48\textwidth]{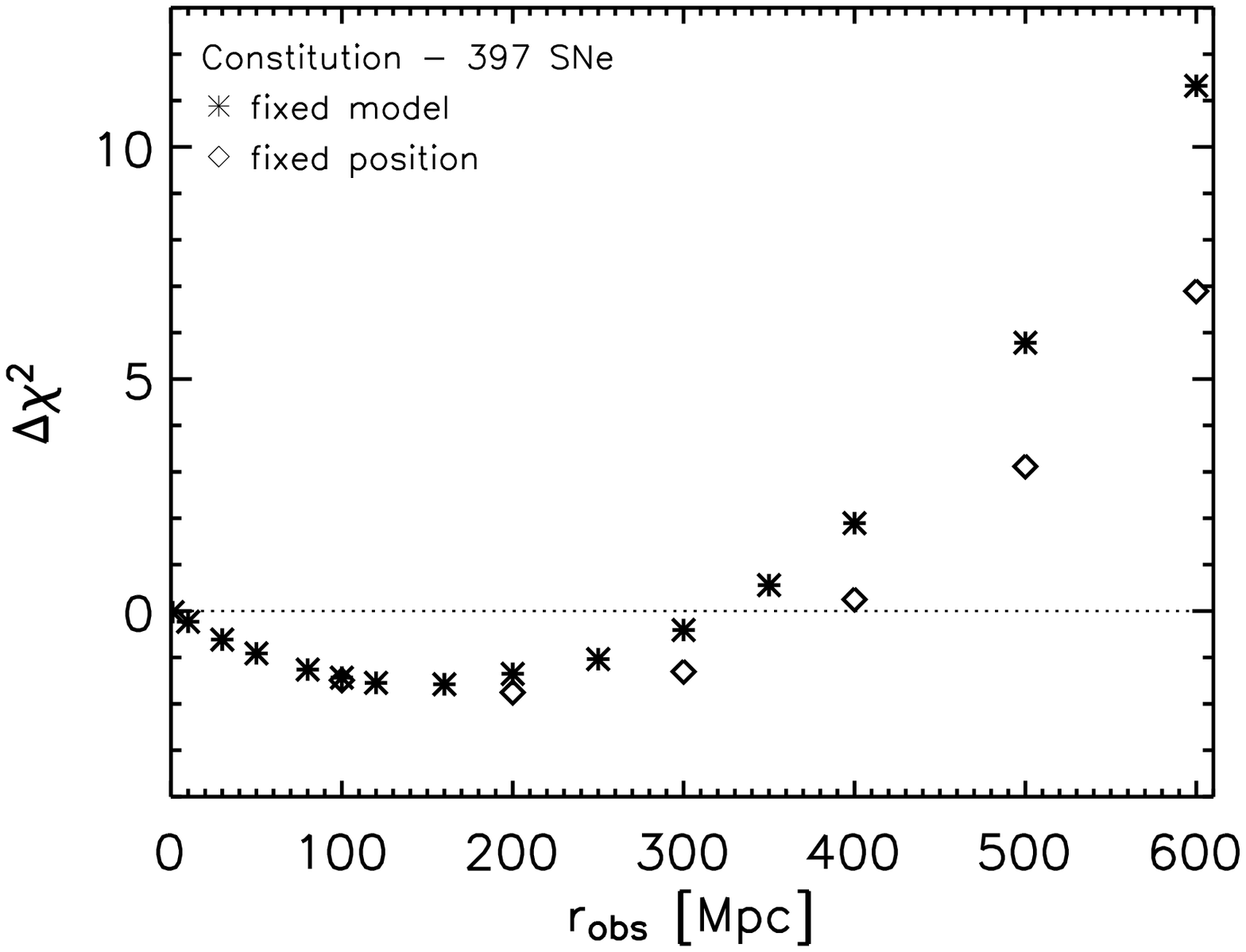}
\caption{\label{fig:chi2} The changes in the $\chi^2$ values 
relative to the on-center value as a function of the observer
position. The stars show the angle-optimized values when the void
model is kept fixed to the best fit on-center parameter values. The
diamonds show the void model-optimized values when the angles are kept
fixed to the angle-optimized values. The scale radius of the void is $r_{\rm
s}=0.7$~Gpc for the SDSS-II set (left panel) and $r_{\rm s}=3.5$~Gpc
for the Constitution set (right panel).}
\end{center}
\end{figure}

We displace the observer from the center and make new fits to the data
sets. Figure~\ref{fig:chi2} shows the changes in the $\chi^2$ values
compared to the on-center value, as a function of the observer
position. In the first case (denoted as fixed model) we take the best
fit on-center model and displace the observer in different directions,
i.e., we scan the parameters $r_{\rm obs}$, $\beta$ and $\sigma$ for a
fixed void model. The stars show the lowest $\chi^2$ value obtained
for each value of $r_{\rm obs}$. Note that the scale radius of the void is
very different for the two data sets, with $r_{\rm s}=0.7$~Gpc for the
SDSS-II set and $r_{\rm s}=3.5$~Gpc for the Constitution set. For the
SDSS-II (left panel) we find that the fit improves over the on-center
case out to about $r_{\rm obs}=250$~Mpc, which corresponds to 36\,\% of
the scale radius. The best fit has $\Delta\chi^2=-0.5$ for $r_{\rm
obs}=100$~Mpc and the void center at $(\alpha _{\rm c},\delta _{\rm
c})=(200^{\circ},-50^{\circ})$ in equatorial coordinates or $(l_{\rm
c},b_{\rm c})=(308^{\circ},13^{\circ})$ in galactic coordinates. For
the Constitution set (right panel) we find that the fit improves over
the on-center case out to about $r_{\rm obs}=320$~Mpc, which
corresponds to 9\,\% of the scale radius. The best fit has
$\Delta\chi^2=-1.6$ for $r_{\rm obs}=160$~Mpc and the void center at
$(\alpha _{\rm c},\delta _{\rm c})=(350^{\circ},-50^{\circ})$ in
equatorial coordinates or $(l_{\rm c},b_{\rm
c})=(334^{\circ},-61^{\circ})$ in galactic coordinates.

As presented in section~\ref{oncenter}, the best fit to the SDSS-II
data for an on-center observer in an asymptotically flat model is a
huge void with $r_{\rm s}=4.8$~Gpc that has $\Delta\chi^2=5.3$
compared to the non-flat model. If we displace the observer in such a
model, the best fit instead has $\Delta\chi^2=0.3$ for $r_{\rm
obs}=600$~Mpc (13\,\% of the scale radius). A model with an off-center
observer in an asymptotically flat void can thus provide a fit that is
comparable to that of an on-center observer in a non-flat void. Note,
however, that this requires three additional parameters.

In the second case (denoted as fixed position) we use the orientation
angles that gave the lowest $\chi^2$ values in the previous case and
instead optimize the void model, i.e., we scan the parameters
$\Omega_{\rm in}$, $\Omega_{\rm out}$ and $r_{\rm s}$ for a fixed
position. Using this approach it is possible to improve the fits
further. For the SDSS-II (left panel of Figure~\ref{fig:chi2}), the
fit is improved all the way out to $r_{\rm obs}\sim 550$~Mpc. As the
observer is displaced further and further from the center, the best
fit parameters change to lower values for $\Omega_{\rm in}$ and
$\Omega_{\rm out}$, and larger values for $r_{\rm s}$. For the
Constitution set (right panel) the fit is improved out to $r_{\rm
obs}\sim 400$~Mpc. The changes in $\Omega_{\rm in}$ and $r_{\rm s}$
are only small and the model remains asymptotically flat.

We would like to point out that we have not performed a full
high-resolution parameter scan over all six parameters simultaneously,
so we expect that it is possible to find (slighty) lower $\chi^2$
values than what we have presented here and we refrain from giving
precise quantitative limits on how far we can be from the center at a
certain level of confidence. Regardless, our results show that SN~Ia
data allow for off-center observers and that the fit in fact can be improved for such a
scenario. For the Constitution set the observer can be displaced $\sim$15\,\%
of the scale radius from the center and still yield an acceptable fit to the data.
For the SDSS-II set a tolerable fit can be obtained for observers displaced all the way out to
around the scale radius. However, given that three additional parameters for the
observer position have been introduced, without providing a
substantial improvement of the fit, we cannot claim that the
off-center model is preferred from a model selection point of view.

\section{Constraining the observer position with SNe~Ia and the CMB dipole}\label{cmb}
Being situated away from the center of the void induces anisotropies
in the CMB temperature~\cite{2006PhRvD..74j3520A}. So far we have
disregarded this effect in the analysis and focused purely on the
SN~Ia data. In this section we will continue to investigate the
constraints on the observer position coming from the SNe~Ia
observations while simultaneously accommodate the CMB dipole
anisotropy measurement. This can be achieved by introducing a peculiar
velocity of the observer directed to counterbalance the dipole induced
by the off-center position. Here we will impose that the void model is
asymptotically flat, $\Omega_{\rm out}=1$, in order to be consistent
with constraints on the spatial curvature derived from measurements of
the CMB.

\subsection{CMB dipole and peculiar velocity}
The COBE satellite measured the average CMB temperature across the sky
to be $T_0=2.725$~K and the amplitude of the temperature fluctuations
to $\Delta T=3.353$~mK~\cite{1996ApJ...464L...1B}. The main
contribution to the CMB temperature anisotropies comes from the
dipole, which is in the direction
$(l,b)=(264.26^{\circ},48.22^{\circ})$ in galactic coordinates or
$(\alpha,\delta)=(168.05^{\circ},-7.06^{\circ})$ in equatorial
coordinates. In a homogeneous universe, the CMB dipole is attributed
to a velocity, $v_{\rm d}$, of the observer relative to the comoving
coordinates,
\begin{equation}
\frac{v_{\rm d}}{c}=\frac{\Delta T}{T_0}\ .
\end{equation}
The COBE measurements imply that $v_{\rm d}=369$~km~s$^{-1}$ (corresponding to the net velocity of the Sun relative to the CMB). The
measured redshifts of the SNe~Ia have been corrected for this velocity
in the data sets, so that they are given in a frame at rest with
respect to the CMB.

In the LTB scenario, the inhomogeneous expansion can be interpreted to
cause a velocity, $v_{\rm h}$, of the off-center observer relative to
the void center~\cite{2006PhRvD..74j3520A},
\begin{equation}
\frac{v_{\rm h}}{c}=\frac{H_{0,\rm in}-H_{0,\rm out}}{c}r_{\rm obs}\ .
\end{equation}
In this picture, it is thus the observer's position in an
inhomogeneous universe, and not primarily a coordinate velocity, that
induces the measured CMB dipole. The CMB photons arriving from the far
side of the void will travel longer through a region with larger
expansion rate and will thus be redshifted. This means that the CMB
temperature measured in the direction of the void center will be lower
compared to that measured in the opposite direction. The void center
is then located at $(\alpha _{\rm c},\delta _{\rm
c})=(348.05^{\circ},7.06^{\circ})$ and the orientation angles are
$\beta = 82.94^{\circ}$ and $\sigma = 11.95^{\circ}$.

It is possible to introduce a peculiar velocity, $v_{\rm p}$, of the
observer, directed towards the void center, as a counterbalance in
order to allow for larger off-center displacements without violating
the CMB dipole measurement. The requirement is thus that it is the net
effect of the off-center position and the peculiar velocity that gives
rise to the observed dipole,
\begin{equation}\label{requirement}
\vec{v}_{\rm h}+\vec{v}_{\rm p}=\vec{v}_{\rm d}\ .
\end{equation}

The introduction of a peculiar velocity, directed along the z-axis,
will also affect the SN~Ia measurements. For a given SN~Ia, the
measured redshift $z$ and luminosity distance $d_{\rm L}$ have to be
translated into their cosmological counterparts $\bar{z}$ and
$\bar{d_{\rm L}}$ according to~\cite{2007ApJ...661..650H}
\begin{equation}\label{zbar}
1+z=(1-v_{\rm p}\cos\gamma)(1+\bar{z})\ ,
\end{equation}
\begin{equation}\label{dlbar}
d_{\rm L}=\bar{d_{\rm L}}(\bar{z})(1+v_{\rm p}\cos\gamma)(1-v_{\rm
p}\cos\gamma)^2\ .
\end{equation}
We have neglected the peculiar velocities of the SNe~Ia. In the
fitting procedure we must thus take the measured redshift and obtain
the cosmological redshift using equation~(\ref{zbar}), calculate the
corresponding cosmological luminosity distance, and finally translate
this value according to equation~(\ref{dlbar}) in order to compare to
the measured luminosity distance.

\subsection{Results}

\begin{figure}
\begin{center}
\includegraphics[angle=0,width=.48\textwidth]{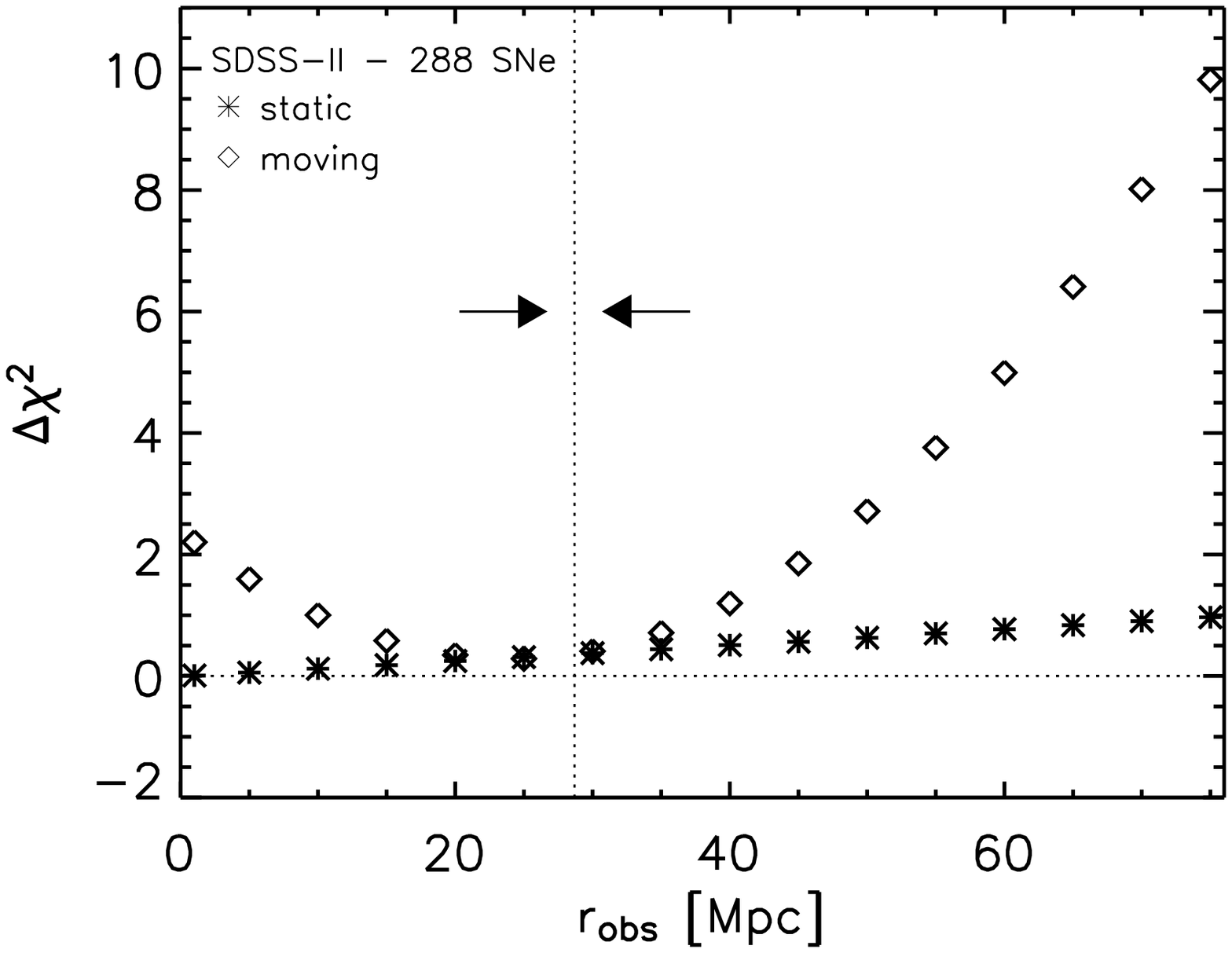}
\includegraphics[angle=0,width=.48\textwidth]{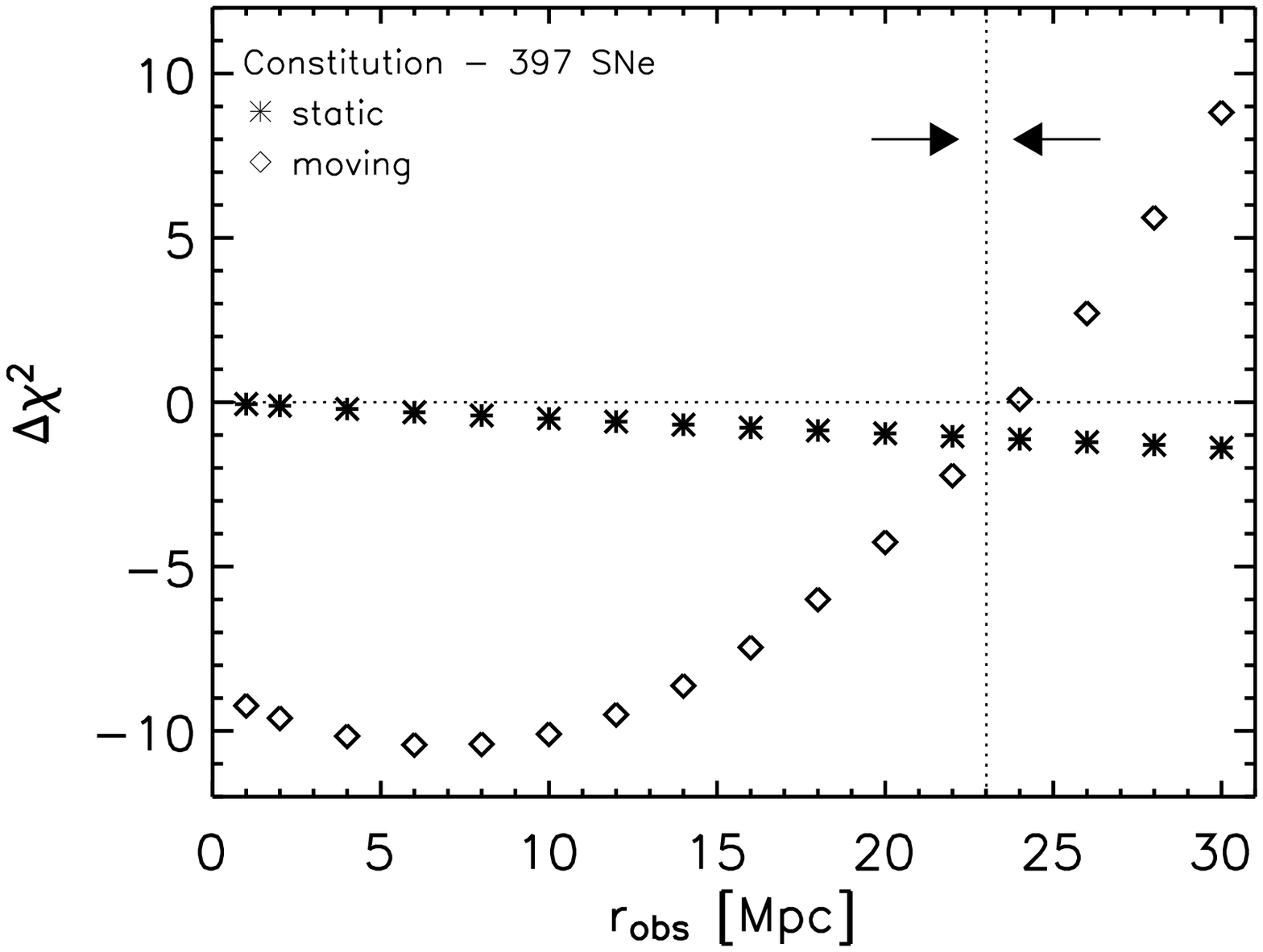}
\caption{\label{fig:chi2_cmb} The changes in the $\chi^2$ values for the fit 
to the SNe~Ia as a function of the observer's position. The stars show
the values when the static observer is displaced in the direction of
the CMB dipole in the best fit on-center LTB model. The diamonds show
the values when the observer also has a peculiar velocity directed to
accommodate the observed CMB dipole. The arrows indicate the direction
of motion, either away from the void center or towards it. The
vertical dotted line shows the position where the peculiar velocity is
zero. The scale radius of the void is $r_{\rm
s}=5.0$~Gpc for the SDSS-II set (left panel) and $r_{\rm s}=3.7$~Gpc
for the Constitution set (right panel).}
\end{center}
\end{figure}

We make new model fits for the case of a static on-center observer in
an asymptotically flat void, using the measured redshifts instead of
the redshifts corrected to the CMB frame provided in the data set. The
best fit to the SDSS-II set has $\chi^2=231.5$ and the best fit
parameters are $\Omega_{\rm in}=0.22$ and $r_{\rm s}=5.0$~Gpc. For the
Constitution set, the best fit has $\chi^2=471.5$, with $\Omega_{\rm
in}=0.13$ and $r_{\rm s}=3.7$~Gpc.

Using these best fit on-center models, Figure \ref{fig:chi2_cmb} shows
how the $\chi^2$ values change compared to the on-center value as the
observer is displaced from the center in the direction of the CMB
dipole. In the first case (denoted as static) the observer has no
peculiar velocity and the CMB dipole requirement is disregarded. These
points only serve as a comparison in the plot. In the second case
(denoted as moving) the observer has a peculiar velocity that
perfectly balances the off-center position so that the dipole
requirement of equation~(\ref{requirement}) is fulfilled at all values
of $r_{\rm obs}$. The arrows indicate the direction of the velocity,
either away from the void center or towards it. The vertical dotted
line marks the position where the peculiar velocity is zero. Figure
\ref{fig:chi2_cmb} demonstrates the power of combining the SNe~Ia data
with the CMB dipole requirement when constraining the observer
position. While the induced CMB dipole can always be balanced with an
appropriately directed peculiar velocity, this motion will
simultaneously affect the quality of the fit to the SNe~Ia. For the
SDSS-II set (left panel) the peculiar velocity is zero at $r_{\rm
obs}=28.7$~Mpc. Introducing a peculiar velocity leads to a
deterioration of the fit compared to the static case. The best fit is
obtained for $r_{\rm obs}=25$~Mpc, which corresponds to 0.5\,\% of the
scale radius. The observer thus has a small peculiar velocity directed
away from the void center for this fit. While smaller displacements
can give comparable fits, the $\chi^2$ value quickly increases as the
observer is displaced further from the center. For the Constitution
set (right panel) the peculiar velocity is zero at $r_{\rm
obs}=23$~Mpc. The peculiar velocity leads to an improvement of the fit
for smaller displacements compared to the static case. The best fit is
obtained for $r_{\rm obs}=6$~Mpc, which is only 0.2\,\% of the scale
radius. The observer has a sizeable peculiar velocity directed away from
the void center for this fit. The $\chi^2$ value increases quickly as
the observer is displaced further from the center.

We conclude that in order to obtain a good fit to the SN~Ia data and
simultaneously accommodate the CMB dipole requirement, the observer
must be located within $\sim$1\,\% of the scale radius.
 
\section{Conclusions}\label{conclusions}
We have considered off-center observers in a large, spherically
symmetric local void described by the LTB metric and investigated the
constraints on the observer position placed by SNe~Ia. The analysis
was performed using two supernova data sets, the first-year SDSS-II
data set and the Constitution data set, which differ in the redshift
distribution and the number of SNe~Ia, as well as the light-curve
fitter used in order to obtain the peak magnitudes. Models with an
on-center observer were able to provide good fits which had slightly
lower $\chi^2$ values than those obtained for the flat $\Lambda$CDM
model, but the best fit voids look very different for the two
data sets. Whereas the SDSS-II data prefer a moderately large void and
a low value of the asymptotic matter density, the Constitution set is
best fit with an asymptotically flat model with a very large void. By
displacing the observer from the center we found that the data indeed
allow for off-center observers. For the SDSS-II set the fit was
improved out to 36\,\% of the scale radius and for the Constitution
set out to 9\,\%, the stronger constraint placed by the Constitution
set as expected from the larger number of SNe~Ia and the more
isotropic sky distribution. However, for both data sets the
improvement of the fit was only marginal. Using SN~Ia data alone, we
conclude that the observer can be displaced at least 15\,\% of the void scale radius from the center and still give an acceptable
fit to the data. These conclusions are in good agreement with those
reached in Alnes \& Amarzguioui~\cite{2007PhRvD..75b3506A}. We have also performed the analysis using a different parameterisation of the density profile 
\cite{2008JCAP...04..003G} with very similar results.

In the final part of the analysis we also took into consideration that
the off-center position induces anisotropies in the CMB
temperature. While the requirement that the induced CMB dipole must be
consistent with the measured value can always be accommodated by
introducing a balancing peculiar velocity of the observer, such a
motion simultaneously affects the SN~Ia observations. Using this
combination of the CMB dipole measurement and the SNe~Ia data, we were
able to determine very strict constraints on how far from the center
the observer can be located. The best fits were obtained for an
observer located at 0.5\,\% of the scale radius for the SDSS-II data set
and 0.2\,\% of the scale radius for the Constitution data set. In order
to still get a good fit to the SN~Ia data, the observer must be
within about 1\,\% of the scale radius.

Our more general conclusion is that within the void model, observers
have to live very close to the center in order for the model to
accomodate the data. Besides requiring an uncomfortable amount of
fine-tuning, such a scenario also constitutes a severe challenge to
the Copernican principle that we should not occupy a special place in
the universe.

In the future, we expect new and better data from many independent
cosmological probes to put the LTB models to even greater challenges,
including new model constraints from large SN~Ia data sets with
extensive sky coverage also at higher redshifts. LTB models may
ultimately prove to not be viable cosmological models, but they can at
least serve as a specific set of toy models to gauge the influence of
large-scale matter inhomogeneities on the light propagation. Such
effects are indubitably present as systematic errors in the
observations and it is important to be able to quantify their
significance in order to further advance precision cosmology.

\ack
MB acknowledges support from the HEAC Centre funded by the Swedish
Research Council. EM acknowledges support from the Swedish Research
Council.

\appendix
\section{The geodesic equations}\label{append}

The photon paths are governed by the geodesic equation. Alnes \&
Amarzguioui~\cite{2006PhRvD..74j3520A} derived the differential
equations for the time coordinate $t$, radial coordinate $r$ and polar
angle $\theta$ in the LTB space-time,
\begin{equation}\label{geot}
\frac{d^2t}{d\lambda^2}+\frac{A'\dot{A}'}{1-k}
\Bigg( \frac{dr}{d\lambda} \Bigg)^2+A\dot{A}\Bigg( \frac{d\theta}{d\lambda} \Bigg)^2=0\ ,
\end{equation}
\begin{equation}\label{geor}
\frac{d^2r}{d\lambda^2}+\Bigg( \frac{A''}{A'}+\frac{k'}{2-2k} \Bigg)
\Bigg( \frac{dr}{d\lambda} \Bigg)^2+\frac{2\dot{A}'}{A'}\frac{dr}{d\lambda}
\frac{dt}{d\lambda}-\frac{A(1-k)}{A'}\Bigg( \frac{d\theta}{d\lambda} \Bigg)^2=0\ ,
\end{equation}
\begin{equation}\label{geoth}
\frac{d}{d\lambda}\Bigg( A^2\frac{d\theta}{d\lambda} \Bigg)\equiv \frac{d}{d\lambda}J=0\ ,
\end{equation}
where the paths are parameterized by the affine parameter $\lambda$
and equation~(\ref{geoth}) has been written as conservation of angular
momentum $J$. Note that due to axial symmetry, the photon paths are
independent of the azimuth angle $\phi$. Furthermore, the equation for
the redshift $z$ reads as
\begin{equation}
\frac{dz}{d\lambda}=-(1+z)\frac{d\lambda}{dt} \Bigg[ \frac{A'\dot{A}'}{1-k}
\Bigg( \frac{dr}{d\lambda} \Bigg)^2+A\dot{A}\Bigg( \frac{d\theta}{d\lambda} \Bigg)^2 \Bigg]\ .
\end{equation}
By introducing $u\equiv dt/d\lambda$ and $p\equiv dr/d\lambda$, we can
bring equations~(\ref{geot}) and (\ref{geor}) down to first order, and the
set of differential equations can then be written as
\begin{equation}\label{dudl}
\frac{du}{d\lambda}=-\Bigg[ \frac{A'\dot{A}'}{1-k}p^2+A\dot{A}\Bigg( \frac{J}{A^2} \Bigg)^2\Bigg]\ ,
\end{equation}
\begin{equation}
\frac{dp}{d\lambda}=\frac{A(1-k)}{A'}\Bigg( \frac{J}{A^2} \Bigg)^2-\frac{2\dot{A}'}{A'}pu-
\Bigg( \frac{A''}{A'}+\frac{k'}{2-2k} \Bigg)p^2\ ,
\end{equation}
\begin{equation}
\frac{d\theta}{d\lambda}=\frac{J}{A^2}\ ,
\end{equation}
\begin{equation}\label{dzdl}
\frac{dz}{d\lambda}=-(1+z)\frac{1}{u} \Bigg[ \frac{A'\dot{A}'}{1-k}p^2+A\dot{A}
\Bigg( \frac{J}{A^2} \Bigg)^2 \Bigg]\ .
\end{equation}
Combining Eqs. (\ref{dudl}) and (\ref{dzdl}) yields
\begin{equation}
u=u_{0}(1+z)\ ,
\end{equation}
where we can choose $u_{0}=u(\lambda=0)=-1$.

We want to solve for the coordinates as a function of redshift, and
this can be achieved since
\begin{equation}
\frac{d}{d\lambda}=\frac{dz}{d\lambda}\frac{d}{dz}\equiv q\frac{d}{dz}\ .
\end{equation}
We can now write down the system of differential equations that we
need to solve to obtain $t(z)$, $r(z)$, $\theta(z)$ and $p(z)$,
\begin{equation}
\frac{dt}{dz}=\frac{u_{0}(1+z)}{q}\ ,
\end{equation}
\begin{equation}
\frac{dr}{dz}=\frac{p}{q}\ ,
\end{equation}
\begin{equation}
\frac{d\theta}{dz}=\frac{J}{qA^2}\ ,
\end{equation}
\begin{equation}
\frac{dp}{dz}=\frac{1}{q}\Bigg[ \frac{(1-k)}{A'}\frac{J^2}{A^3}-\frac{2\dot{A}'}{A'}pu_{0}(1+z)-
\Bigg( \frac{A''}{A'}+\frac{k'}{2-2k} \Bigg)p^2\Bigg]\ ,
\end{equation}
under the constraint
\begin{equation}
q\equiv \frac{dz}{d\lambda}=-\frac{1}{u_{0}} \Bigg[
\frac{A'\dot{A}'}{1-k}p^2+ \frac{\dot{A}J^2}{A^3}\Bigg]\ .
\end{equation}

\section*{References}

\end{document}